# The energy landscape of DNA-binding proteins along the genome

*Alessandro Pandolfi[1], Riccardo Beccaria[2,3]* and Guido Tiana[4]**
[1]Istituto Europeo di Oncologia, via Adamello 16, 20139 Milano, Italy
[2]Molecular and Cellular Modeling Group, Heidelberg Institute for Theoretical Studies (HITS), 69118 Heidelberg, Germany
[3]Faculty of Physics, Heidelberg University, Im Neuenheimer Feld 227, 69120 Heidelberg, Germany
[4]Department of Physics, Università degli Studi di Milano and INFN, via Celoria 16, 20133 Milano, Italy

*Correspondence to:
Riccardo Beccaria, tel. +49 (0)6221533259, email: Riccardo.Beccaria@h-its.org;
Guido Tiana, Department of Physics, Università degli Studi di Milano, vai Celoria 16, 20133 Milano, Italy, tel. +39-0250317221, email: guido.tiana@unimi.it

ABSTRACT
Reconstructing the energy profile of DNA-binding proteins along the genome requires an algorithm that quantifies efficiently the binding free energy. We assembled a dataset of protein structures and DNA binding sites, together with their binding energies, and used it to train a machine-learning algorithm that learns a latent invariant representation of the protein interface and of the DNA sequence, combining them to predict the free energy. After validating the method, we used it to determine the energy profile of a single-domain transcription factor (PU.1) sliding on mammalian chromosomes, predicting its binding regions and quantifying the statistical properties that determine their stability and their kinetic accessibility.

## INTRODUCTION

The energetics of protein-DNA interactions play a key role in determining the binding sites and the dynamics of transcription factors, histones, replication proteins, chromatin remodelers and many other proteins that control the cellular behavior. The energy landscape of proteins sliding over chromosomes determines the thermodynamic and the kinetic properties of the protein-DNA interaction, determining the probability and the timing of critical cellular processes, such as gene activation and repression. Quantifying this is therefore essential for understanding the physical mechanisms of transcription.

Several algorithms have been developed to predict how proteins bind to specific DNA sequences *ab initio*, that is without the need of high-throughput experiments done on the specific system under study. A standard way of classifying DNA binding sites for a given protein is to score them using Position Weight Matrices (PWM), which are obtained from the empirical frequencies of available complexes[1]. Machine learning methods, either shallow as random forests[2] or deep as the convolutional neural network of DeepBind[3] or DeepSea[4], show improved classification abilities than PWM. A careful description of the molecular geometry of the complex was found to improve the classification[5]. However, the output of these algorithms is a confidence score, not a binding free energy. Even in the case of the transformer-based architecture of Enformer[6], which is trained on cellular data such as ChIP-seq and CAGE tracks, the output is a predicted read count, which are only indirectly related to the free-energy binding between proteins and DNA.

Poisson-Boltzmann calculations of the electrostatic energy, summed to a knowledge-based term, was used to simulate the dynamics of the complex and to calculate its free energy[7]. The Rosetta force field was successfully used to redesign a DNA-binding protein, optimizing the binding energy[8]. FoldX also predicts realistic binding free energies by combining a statistical force field with local structure minimization[9]. An important finding is that the performance of these algorithms, particularly those based on a statistical potential, depend critically on the atomic-level description of the system[10]. Anyway, they are usually fast enough to predict the energetic effects of mutations or to optimize the interactions binding complex. However, they are too slow for the more ambitious goal of scanning the energy landscape of DNA-binding proteins over large DNA sequences or on whole genomes.

Our first step to study such a landscape was the development of a machine learning (ML) model to quantify the binding free energy given the atomic-scale structure of the protein and the sequence of a DNA region. We encoded separately the binding region of the protein and the binding site in DNA into two abstract representations that are invariant for roto-translations and for indexing of the atoms, combining them only at the end to predict the binding free energy. We built a dataset of protein-DNA binding free energies, combining binding annotations, protein structures and thermodynamic data concerning binding free energies, obtained from different databases. The dataset was used to train a ML model that generates the abstract representation of the molecular pairs and hence the binding free energy.

The evaluation of the free energy for a protein whose binding motif is known to DNA is sufficiently rapid to permit its traversal across entire chromosomes of a genome. We did it for protein PU.1, a transcription factor that binds a contiguous DNA sequence[11]. In this way, we can obtain the energetic binding profile of the protein and study its statistical properties.

## METHODS

*Creation of a dataset of protein-DNA complexes and their binding free energies*

Protein and DNA annotations were retrieved from the Nucleic Acid Knowledge Base (NAKB), three-dimensional structural data from the Protein Data Bank (PDB), and thermodynamic experimental data from the ProNAB database[12]

The ProNAB database contains data and metadata regarding binding affinity experiments of protein-nucleic acid complexes. We then queried all PDB biological assemblies that contain any protein found in the ProNAB database, identified by their UniProt id. From the biological assemblies we extracted structural and sequence-derived annotations, such as protein binding domains, protein and nucleic acid binding sequences, interface amino acids and nucleotides, nucleic acid type, strandedness, and other structural descriptors. Because a substantial fraction of ProNAB entries corresponds to proteins containing experimentally introduced mutations, the corresponding amino acid substitutions were introduced *in silico* into the associated PDB structures prior to further structural analysis (see Extended Methods in the SI).

An atomic contact was defined as a pair of atoms separated by no more than *5.0* Å, consistent with the typical range of non-covalent protein–DNA interactions. Interface contacts were defined as atomic contacts involving one protein atom and one nucleic acid atom. The collection of protein and nucleic acid atoms participating in interface contacts was considered the protein–nucleic acid binding interface. All available interfaces were extracted from the PDB biological assemblies previously queried.

By merging information from all data sources, we built an interim dataset composed of triplets of data which will be used by our model, each triplet contains the interface extracted from PDB, the experimental DNA sequence queried from ProNAB and the experimental ΔG measurement queried from ProNAB. Entries were then filtered according to the following criteria: 1) complexes containing canonical double-stranded B-form DNA bound by at least one protein were retained; 2) nucleosome and polymerase complexes were excluded because of their structural binding heterogeneity; 3) interfaces containing fewer than 20 interacting proteic atoms were discarded, as they were considered unlikely to contain sufficient information to characterize protein–DNA binding; 4) entries associated with protein-DNA pairs which have ΔG standard deviation greater or equal than 1 kcal/mol were excluded, as they more likely contain outliers in experimental measurements.

*Machine-learning based potential*

To predict protein–DNA binding affinity ΔG, we developed a multi-modal, asymmetric dual-stream neural network that learns independent latent representations of the protein interface and the DNA sequence before combining them for a final prediction. All the details are described in the Extended Methods in the SI.

The protein binding interface is represented as a spatial geometric graph, following a graph representation inspired by previous work for protein–ligand active decoy classification[13]. Nodes correspond to individual protein atoms and edges connect atom pairs whose Euclidean distance falls within a 5.0 Å cutoff, independently on the structure of covalent bonds. Each atom is embedded from its atomic number via a learnable lookup table, and interatomic distances are expanded into continuous edge features using a set of Gaussian radial basis functions. A SchNet-based continuous-filter graph neural network (GNN) then iteratively refines the atomic embeddings over several interaction blocks, aggregating information from spatial neighbors through continuous-filter convolutions weighted by a smooth cosine cutoff function that damps interactions near the radius boundary. After a residual update, the atomic embeddings are combined via global sum pooling into a single, fixed-size, rotation-invariant latent vector describing the entire binding interface.

In parallel, the DNA sequence is one-hot encoded into a four-channel representation of the canonical nucleotides (A, C, G, T), with shorter sequences zero-padded to a common maximum length. This tensor is processed by a one-dimensional convolutional neural network (CNN) consisting of several convolutional layers, each followed by ReLU activation and batch normalization to stabilize training and extract increasingly contextual sequence features. An adaptive average pooling layer then collapses the sequence dimension into a fixed-length dense embedding, regardless of input sequence length.

The protein and DNA embeddings are concatenated into a joint cross-modal representation and passed through a feed-forward neural network (FFNN) with ELU activations and dropout regularization, culminating in a single scalar estimate of the binding energy. To respect the bidirectional nature of double-stranded DNA, the model is evaluated on both the forward sequence and its reverse complement (obtained by flipping the sequence axis and inverting the nucleotide channels), and the final predicted binding energy is taken as the average of the two resulting predictions.

The complete architecture is trained end-to-end by minimizing the mean squared error between predicted and experimental binding energies over each training batch, with parameters updated using the Adam optimizer. To ensure the model learns genuine binding specificity rather than generic compositional biases, training on native co-crystallized complexes is augmented with a dynamic decoy sampling strategy: each true complex is paired with several synthetic mismatched protein–DNA pairs formed by randomly assigning non-native protein interfaces to DNA sequences. These decoys are treated as weak, non-specific interactions with affinities drawn from a Gaussian distribution (mean −1.0, standard deviation 0.5) and clipped at a lower bound of −2.5 kcal/mol to avoid overlapping with genuine high-affinity binders. To keep the optimization signal balanced, native complexes contribute full weight to the loss while each decoy is downweighted in inverse proportion to the number of decoys per true complex, preventing the large number of synthetic pairs from overwhelming the learning of true, sequence-dependent binding specificity.

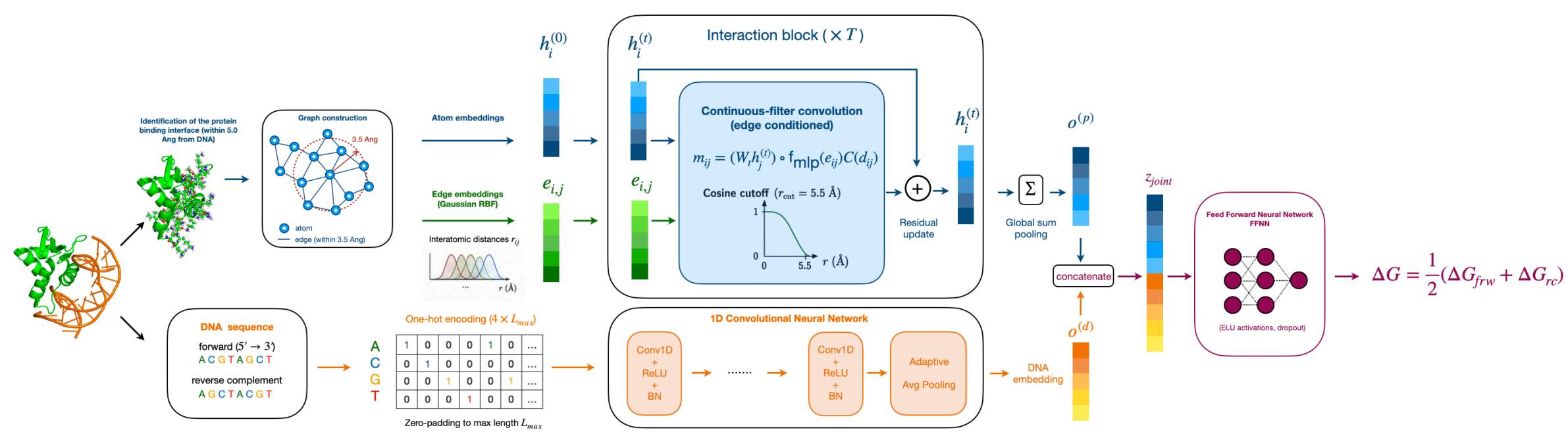

Figure 1: A sketch of the algorithm we used to predict the binding free energy between a protein and a sequence of DNA.

## Results

*The protein-DNA dataset*

We generated the dataset of binding free energies associated with paired protein binding sites and DNA sequences, as described in the Methods. The dataset comprises 151210 entries, then aggregated into 27255 entries by averaging ΔG measurements of identical pairs, finally remaining with 24923 entries after filtering with constraints described in the Methods. The final dataset comprises 266 unique proteins, as identified by their UniProt id, 875 PDB structures and 7312 interfaces. In average, each protein corresponds to 3.3 PDB structures, each PDB structure corresponds to 8.4 interfaces, and each protein corresponds to 27.5 interfaces.

Within the dataset, protein annotations are unbalanced, with some classes being heavily overrepresented, i.e. lambda-like, nucleases, and homeodomains (Fig. 2D). Experimental DNA sequence lengths consistently fall within the 10-70 base pair range, with an average of 26.4 base pairs, a secondary peak at 40 base pairs and an overall right skewed distribution, indicative of variability in experimental designs (Fig. 2B). On the contrary, protein-bound DNA sequences show a narrow normal distribution around an average of 10.5 base pairs, with few outliers that fall in the 20-90 base pair range (Fig. 2C), corroborating previous literature [14,15].

The binding free energies range from -5 to -18 kcal/mol, where typical values for poorly-specific binding proteins is of the order of -8 kcal/mol [16] (Fig. 2E) .

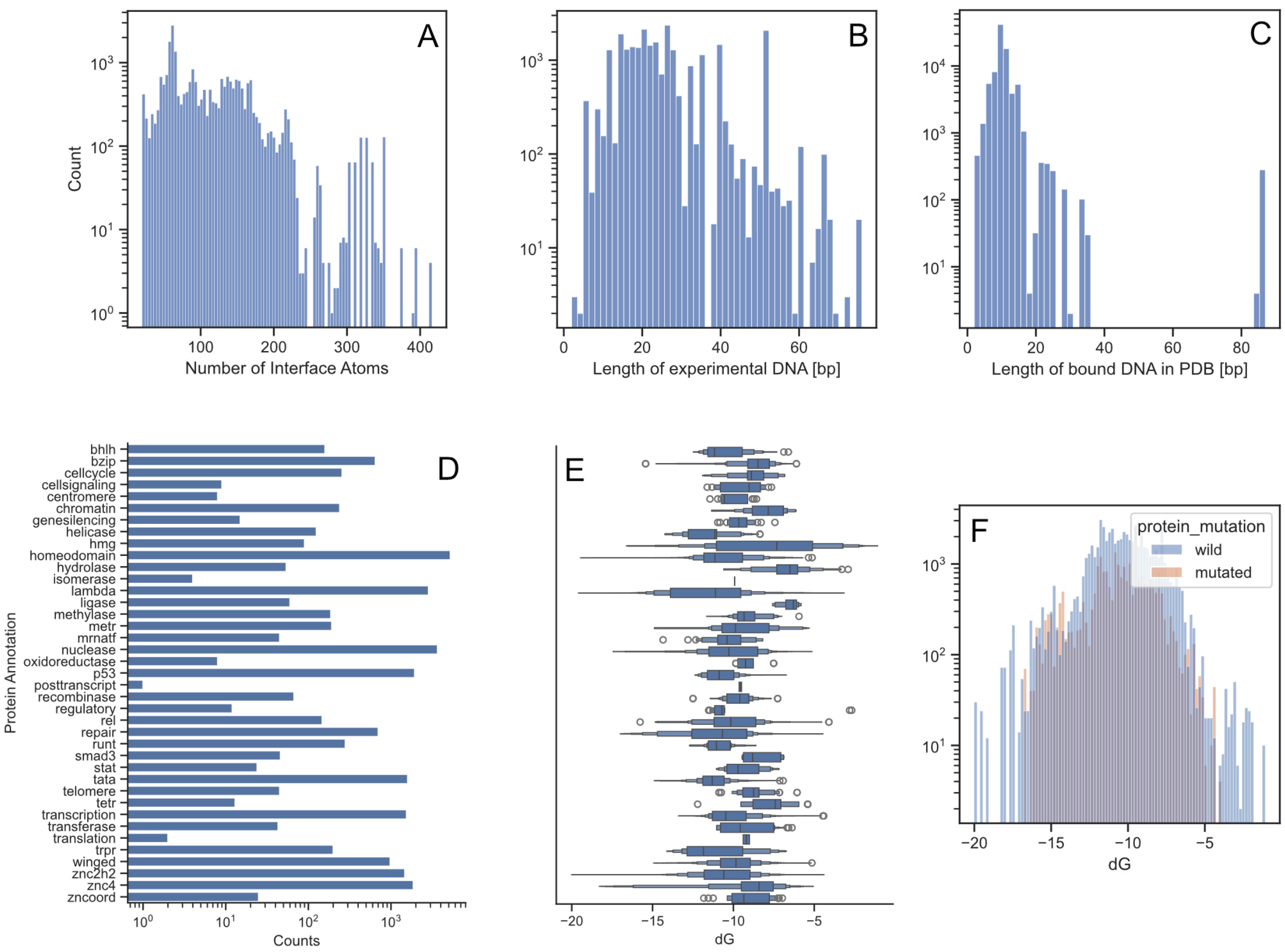

Figure 2: Distributions of relevant variables in the dataset. A) distribution of number of interface atoms; B) length of experimental DNA sequences; C) length of protein-bound DNA sequences extracted from structures (before aggregating); D) distribution of protein annotations; E) ΔG distribution grouped by NAKB protein annotations; F) ΔG distribution grouped by protein wild types and mutated proteins.

*Generation of an accurate machine-learning based potential*

To generate a realistic potential based on the available data, we conducted a comprehensive grid search scanning across a broad hyperparameter space (cf. Table S1 in the SI). In total, several hundred distinct models were trained on an 80% partition of our core dataset across three different data-splitting strategies (detailed below). Initial model selection was guided by performance on an internal validation set comprising the remaining 20% of the data. To robustly assess the generalization capabilities of the top-performing architectures on independent biological contexts, we further benchmarked them against two distinct external challenges. The first one is the prediction of ΔΔG shifts resulting from pointwise mutations in the Lac and Cro repressor operator sites. The second one is the discrimination of native, high-affinity binding genomic sequences identified via ChIP-seq experiments against randomized chromosomal sequences. The optimal hyperparameter configurations for these top-performing models are provided in the SI. Notably, the architecture that yielded the highest accuracy on the ChIP-seq classification task did not coincide with the optimal model for predicting ΔΔG mutational effects (cf. Fig. S1 in the SI). This discrepancy is expected, as the macroscopic task of identifying genome-wide binding profiles against capturing fine-grained energetic fluctuations driven by single-nucleotide variations represents inherently distinct physical and statistical challenges.

For the internal validations, Pearson correlation coefficients r and scatter plots were computed exclusively using native co-crystallized complexes, excluding synthetic decoy pairs (cf. Fig. S2 in the SI for same analysis including decoys), with the goal of assessing the capacity of the model to predict genuine, high-affinity interactions. The first validation strategy was done through a random split of the data (left panel in Fig. 3). As expected, this approach yielded the highest predictive accuracy, achieving a Pearson correlation coefficient of r=0.929. Because this random assignment imposes no constraints on sequence or structural similarity (other than merging identical pairs), closely related data points, such as protein-DNA complexes differing by only a single point mutation, can be shared between the training and validation subsets.

To test the sensitivity of the model to structural variations within the protein interface, we applied a mutation-based split where protein mutations were selected so that the corresponding entries represented approximately 5% of the dataset. All entries associated with these mutations were assigned to the validation set, while the remaining entries, both wild type and mutated, were used for training. Under this scheme (middle panel in Fig. 3), the model achieved a Pearson correlation coefficient of r=0.711. This demonstrates that the network effectively generalizes across distinct protein interfaces containing amino acid substitutions, even when the interface is structurally dissimilar from the training data.

The most stringent validation strategy was done splitting the data based on DNA sequence, where unique DNA sequences from wild type protein entries are randomly sampled and assigned to the validation set (right panel in Fig. 3). This is the most important set for our application, that is calculating the binding energy of a specific protein to uncharacterized genomic regions. Here, the model maintained a good predictive capacity, yielding a Pearson correlation coefficient of r=0.632. Notably, the model accurately captured strong binding affinities approaching −20 kcal/mol, corresponding to the most stable complexes. Moreover, when the prediction is poor, it tends to underestimate the binding free energy. For completeness, we report the Pearson correlation coefficient distributions across the three training schemes for the hundreds of trained models in Fig. S3.

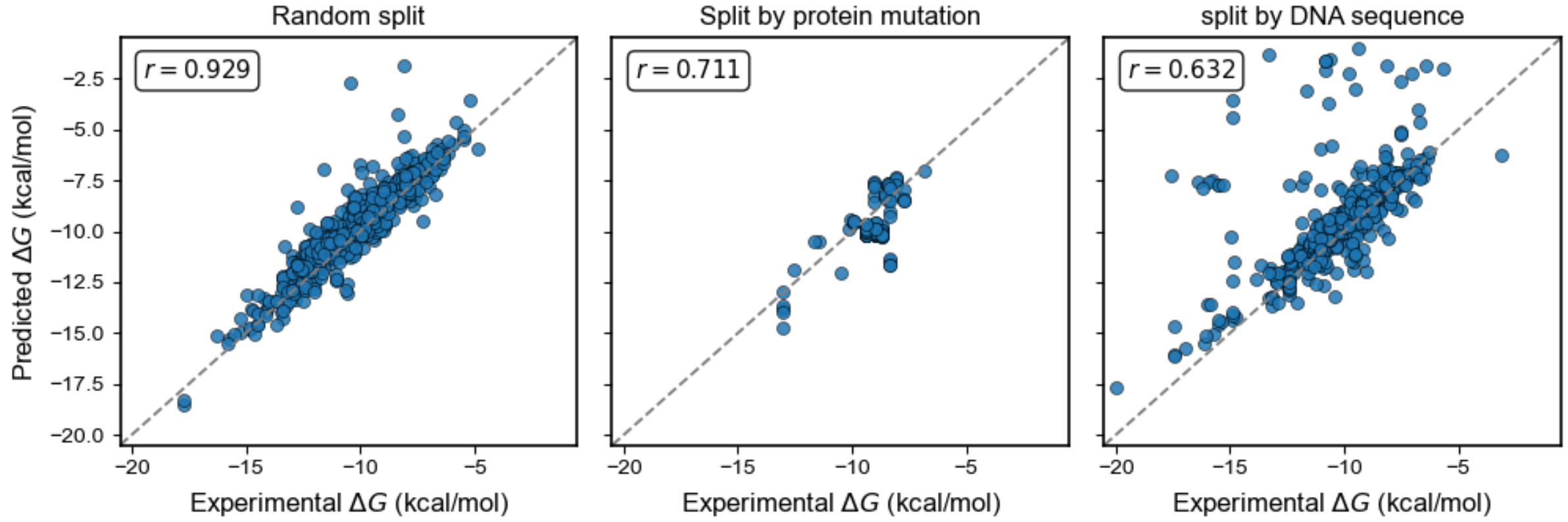

Figure 3: Predicted vs experimental binding free energies using the optimal model for different dataset splitting strategies. For each of them is indicated the Pearson coefficient $r$.

We then tested the ability of the model to predict the binding free energies $\Delta\Delta G$ resulting from single-point mutations in DNA operator sites. We benchmarked our predictions against experimental data characterizing pointwise all base mutations across the binding site of both the l-repressor [17] and Cro repressor[18] proteins. Both proteins bind as dimers the two half-binding sites made overall of the same 17 bases[19].

The prediction model was trained leaving the DNA sequences of the l and Cro operator out of the training set. From all trained models, we selected the two that performed best on the binding free energies ($\Delta\Delta G$) for both repressors (see Fig. S4 for the complete Pearson coefficient distributions of all trained models applied to single-point mutations in DNA operator sites). The correlation found between predicted and experimental $\Delta\Delta G$ of the l-repressor is r=0.62 (top-left panel in Fig. 4), similar to that found in the internal validation with the dataset split by DNA sequence. The correlation coefficient associated with the prediction of the mutations in the binding of Cro repressor is slightly lower, r=0.47, mainly due to some outliers (bottom-left panel in Fig. 4).

To rationalize the errors in the prediction, we inspected the difference $\Delta\Delta G_{pred} - \Delta\Delta G_{true}$ between predicted and true energetic effect of each mutation in the 17 sequence positions for all possible alternative bases (top-right and bottom-right panels in Fig. 3). The sites where the prediction is particularly inaccurate (marked with an asterisk, where the error is greater than 2 kcal/mol) tend to be located at the ends of the two half-sites, farther from the recognizing a-helix of the helix-turn-helix motif. On the other hand, there is not a clear correlation between the prediction error and the degree of conservation of the basis (middle-right panel in Fig 4), that one would observe for a model overfitted on the specific data associated with this system.

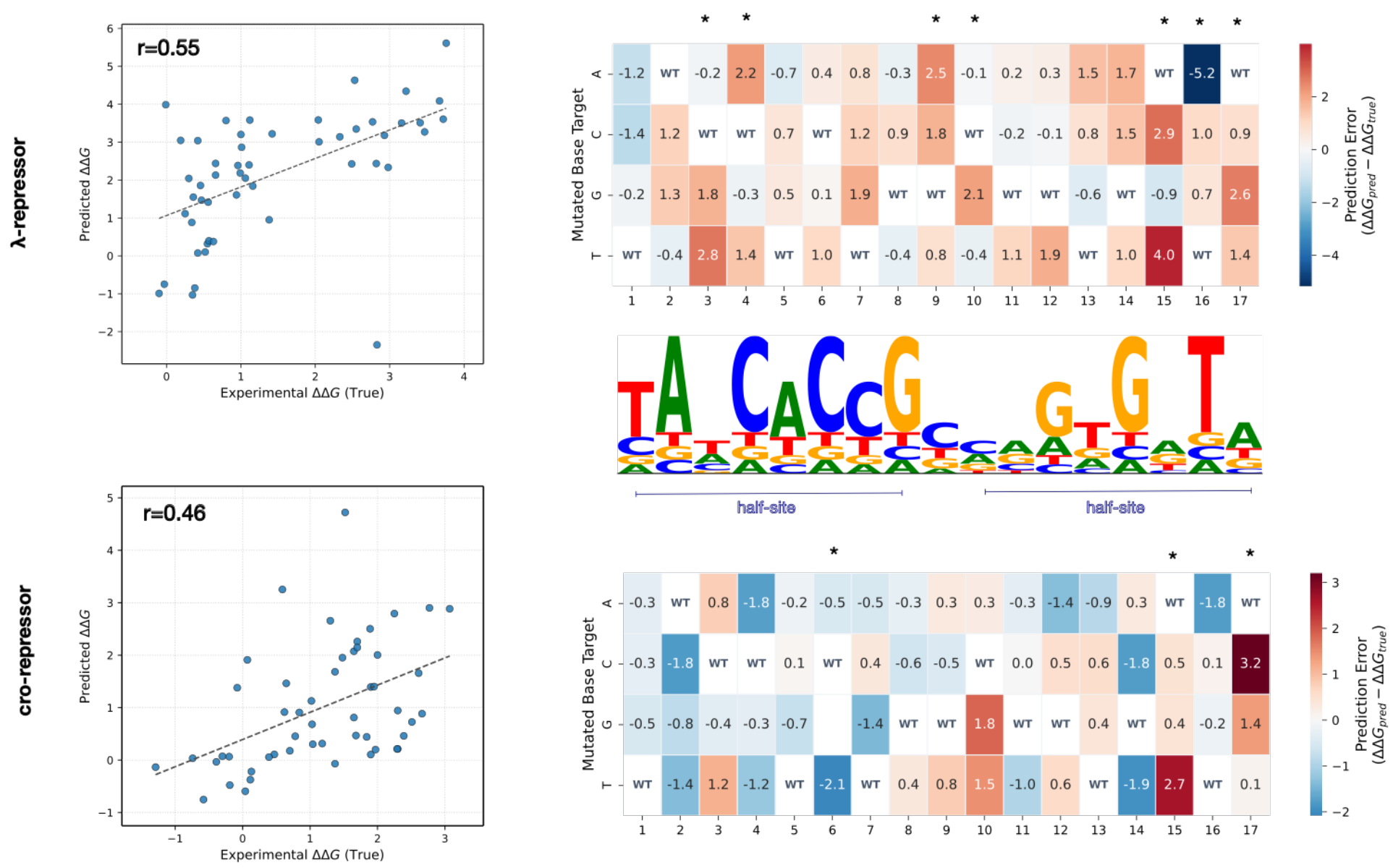

Figure 4: Left panels, the predicted change in free energy $\Delta\Delta G_{\mathrm{pred}}$ versus the experimental one $\Delta\Delta G_{\mathrm{true}}$ for the l-repressor (upper panel) and cro-repressor (lower panel); the dashed line is the least-square fit. Right panels, the positional residual matrices ($\Delta\Delta G_{\mathrm{pred}} - \Delta\Delta G_{\mathrm{true}}$) that describes the prediction error for the two proteins; the sites with errors larger than 2 kcal/mol are marked with stars. Between them, the PWM describing the conservation of bases across the 6 natural

binding sites of the two proteins. The two half-sites where the dimeric repressor bind are indicated below the conservation plot.

*Energy landscape of the genome*

Thanks to its combined efficiency and accuracy, our energy predictor is well-suited to studying the energy landscape of a DNA-binding protein sliding over entire chromosomes. For the sake of simplicity, we studied the murine transcription factor PU.1 (UniProt code P17433) because it recognizes a contiguous DNA region with a single domain (Fig. 5a), thus avoiding the problems associated with tandem binding and bending of the double helix. We focused our attention to chromosome 2 of mouse, which is rather large (182 Mb) and has a balanced mixture of euchromatin and heterochromatin. We evaluated the binding energy on all overlapping windows of 13 bp of DNA, corresponding to the consensus sequence according to the Jaspar database [20].

The typical profile of the binding energy (Fig. 5b) fluctuates approximately from -2 to -12 kcal/mol. The distribution of binding energies across the whole chromosome (Fig. 5c) has a sharp Gaussian-like peak around -2.3 kcal/mol and a fat tail towards negative values. To validate the idea that this tail is generated by genuine binding sites of PU.1, we compared the histogram of energies of 210396 binding sequences determined by chip-seq experiments and extracted from the Jaspar database to that of the same number of sequences, selected randomly on the chromosome. The former histogram peaks at -9 kcal/mol, which is a much lower energy than the -2.5 kcal/mol peak that characterizes sequences of random chromosomal positions (Fig. 5d).

The energies of DNA sites display some degree of correlation as a function of the genomic distance. The autocorrelation function of the energies decays as a double exponential (Fig. 5e), showing two length scales of approximately 3 bp and 75 bp, respectively. We argue that the shorter length scale, corresponding to the major drop in correlations, reflects the resolution of the model; namely, as one slides the scoring window along the DNA sequence, neighboring positions will share a large fraction of their interface atoms and contacted bases. Conversely, the 75 bp length scale is likely to reflect some bulk genome compositional structure, such as GC-content domains, tandem repeats, etc.

The autocorrelation function at longer length scales is dominated by statistical noise. To better investigate the large-scale structure, we inspected the power spectrum of the system, calculated from fast Fourier transform (Fig. 5f). It displays a clear power law with exponent -1.22 in the range of frequencies between $10^{-8}$ to $10^{-5}$ $bp^{-1}$, corresponding to length scales between 100 kbp and 100 Mbp. We notice that this is the range of topological associating domains in mammalian chromatin[21] and that the exponent is comparable to that of several genomic signals previously studied [22–25]. The high-frequency ($> 10^{-5}$ $bp^{-1}$) part of the power spectrum is much flatter and is likely to correspond to the longer correlations observed in the autocorrelation function.

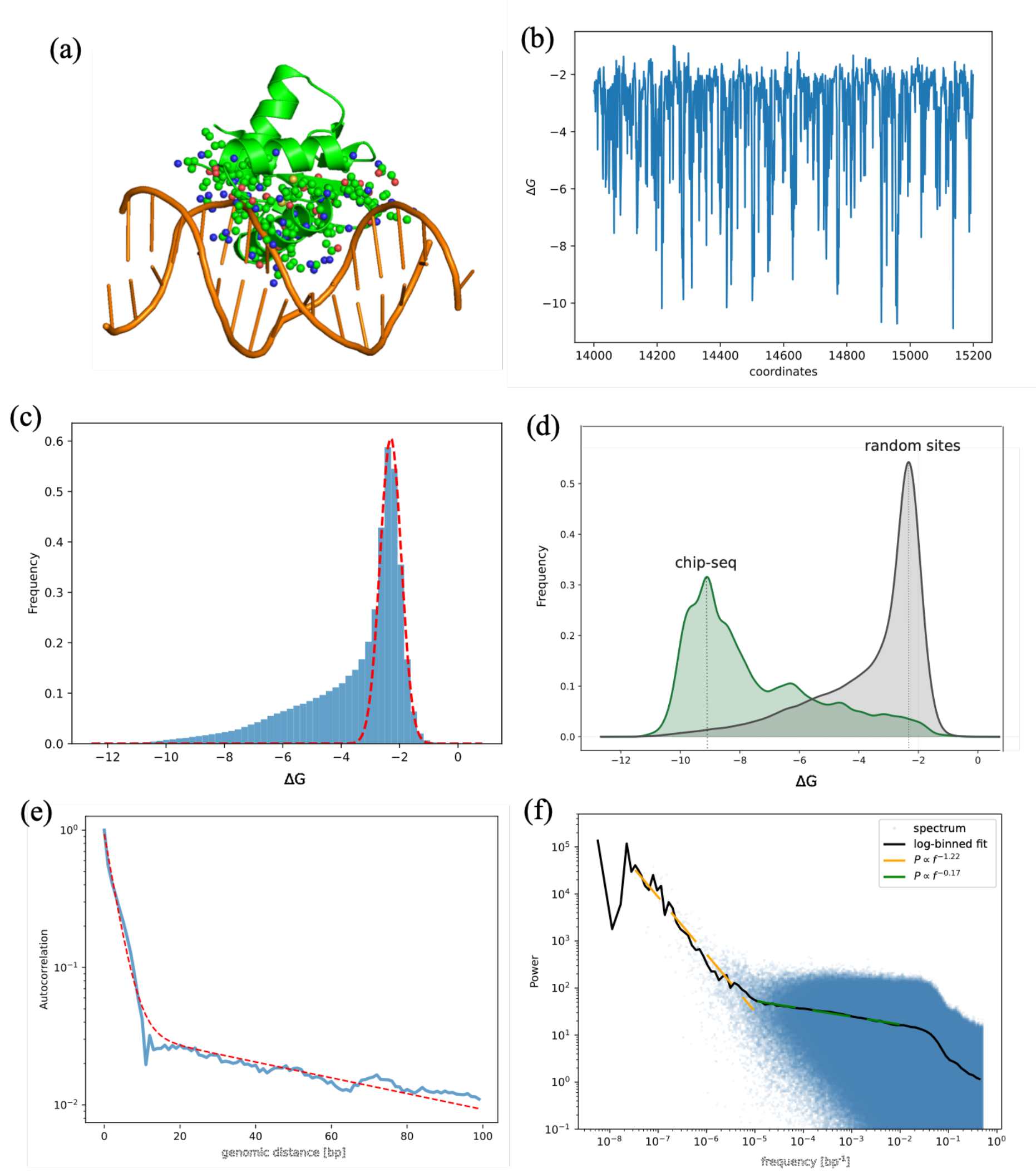

Figure 5: (a) The structure of PU.1 bound to DNA (pdb code: 1PUE, green ribbon). The atoms of the binding pocket are represented with beads. (b) An example of the energy landscape if PU.1 over a region of chromosome 2 of mouse. (c) The distribution of binding energies over the chromosome; the dashed line indicates a Gaussian distribution fitting the high-energy part of the histogram. (d) The distribution of binding energies calculated on the 210396 binding sequences determined by chip-seq and the same number of randomly selected on the chromosome. (e) The autocorrelation function of the binding energy (blue curve) in semilog scale, fitted with a double exponential (dashed red line). (f) The power spectrum of the binding energy (blue dots), their mean in log-spaced bins (black curve) and the power-law fit of two regions of the curve (orange and green dashed lines, respectively).

Besides the general statistical properties of the energy landscape, we can study the sequence similarity characterizing low-energy sites and known binding sites of PU.1, which are in general low-energy states of such a profile (Fig. 5d). The sequences of DNA identified in the Jespar database by chip-seq experiments are very much alike, having a mutual Hamming distance centered around 4 (Fig. 6). Conversely, the whole set of low-energy sequence is much more diverse, displaying a peak at mutual Hamming distance of approximately 10, independently on the energy threshold used to select them. This is also the typical distance between low-energy sequences and experimentally identified binding sites (Fig. S5 in the SI). Thus, experimentally identified sequences are a homogeneous subset of all low-energy ones.

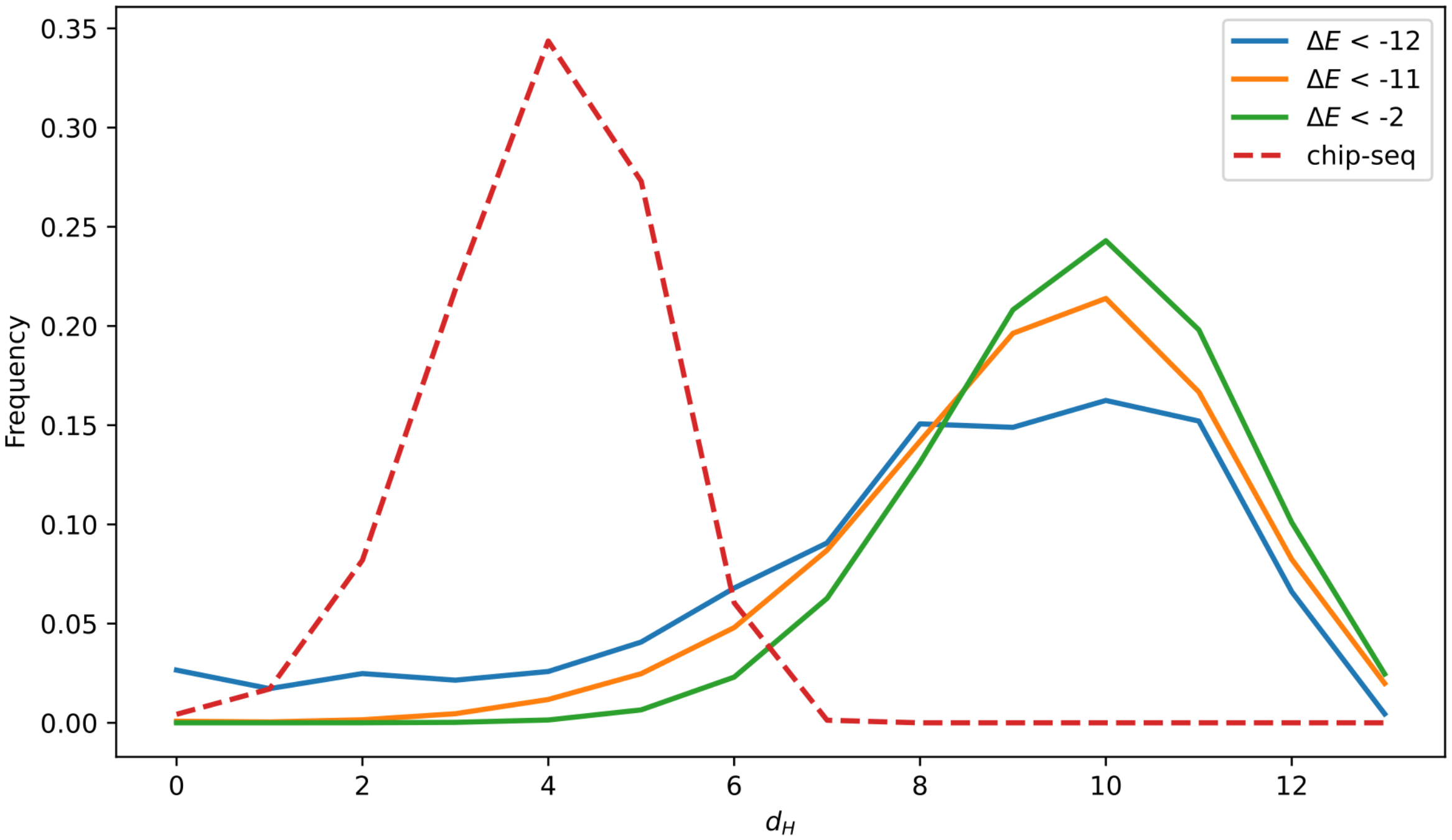

Figure 6: The distribution of Hamming distances between pairs of binding sequences of chromosome 2 with energy below -12 kcal/mol (blue curve), -11 kcal/mol (orange curve), -2 kcal/mol (green curve) and between the sequences obtained in chip-seq experiments (red dashed curve).

One could ask what specific properties distinguish sequences identified by chip-seq among the larger set of low-binding-energy DNA sequences. We noticed that the energy profile around them is somewhat wider than that of low-energy sequences with the same mean energy (Fig 7a). While the width of typical energy minima is of the order of few base pairs, that of experimentally validated ones has a basin spanning approximately 10 bp, giving them a kinetic advantage for binding.

The most conserved bases of the experimentally determined sequences (Fig 7b) set the consensus sequence to AGGAG (defined as the sites with an information content larger than 1 bit). This sequence can be easily aligned with the most likely sequences predicted at low binding energy (Fig. 7c, E<-12 kcal/mol), matching exactly 5 base types out of 6, the missing one matching the next-to most likely. DNA sequences with higher binding energies, even those at the average energy -9 kcal/mol, typical of experimentally determined sequences, display more uniform base probabilities and do not match the consensus sequence (Fig. S6 in the SI). This fact suggests that experimentally determined sequences lie in some of the basin of the lowest energy minima (~-12 kcal/mol), but at higher energy than the minimum.

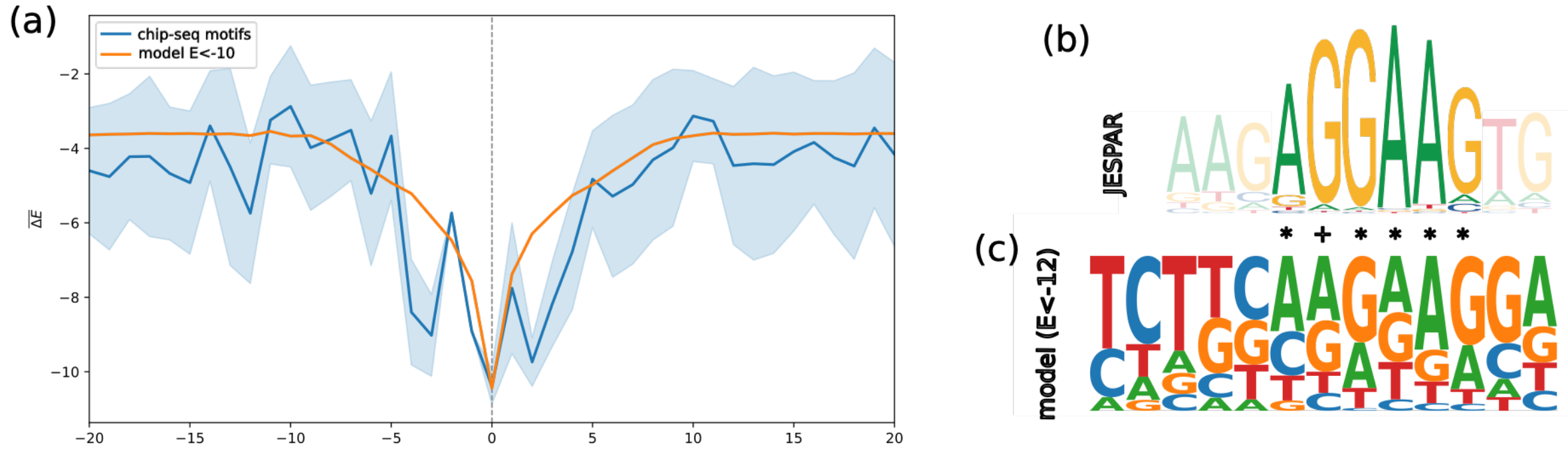

Figure 7: (a) The mean energy profile around 5000 chip-seq sequences (blue curve, the shadow indicating the standard deviation) and the average energy around sequences with $\Delta E < -10$ kcal/mol, having the same average energy (orange curve). (b) The binding profile of PU.1 obtained from chip-seq experiments in terms of PFM. The sites with information content (that is the total height of the symbols) greater than 1 bit are indicated by solid colors. (c) The probability binding profile of the sequences of chromosome 2 with $\Delta E < -12$ kcal/mol. The symbols between the two profiles indicate the best alignment: the asterisk (*) shows that the most probable base of the low-energy sequences is the same as that of the chip-seq sequences; a plus (+) shows that the next-most probable base is optimal.

**DISCUSSION AND CONCLUSIONS**

Understanding how DNA-binding proteins, such as transcription factors and architectural proteins, recognise and bind to their correct targets quickly and stably based on their DNA sequences is a long-standing problem[1,26,27]. The insight we can obtain from cellular experiments is limited to ChIP-Seq or Cut&Tag enrichment, which can identify qualitatively where proteins bind but its affected by multiple factors, such as chromatin accessibility, antibody affinity, etc. Quantitative binding assays can be done with Selex techniques only *in vitro* on a selected set of DNA sequences. In absence of suitable experimental techniques, a fast and reliable predictor of the profile of protein-DNA binding free energy can be a useful tool to elucidate the physics of protein-DNA recognition.

The strategy we followed to build such a predictor is to leaverage on the sizeble, but somewhat unorderd, amount of structural and thermodynamic data available for protein-DNA complexes. Merging the information contained in different databases and pruning them, we obtained a dataset of approximately 25000 complexes. We think that this dataset has an intrinsic value and can be used for several tasks. We used it to train a machine-learning algorithm that predicts the binding free energy from the atomic structure of the binding region of the protein and from the DNA sequence. Using an approach similar to that we used for classifying the binding of small molecules [13], we used as predictor a ML architecture that first generate an abstract representation of the atomic structure of the protein interface and of the DNA sequence, and then put them together to calculate the binding free energy. The representation of the interfacial atoms is invariant for rototranslations and for permutation of their identification index, allowing us to test different selections of the atoms involved in the binding.

We validated the model in different ways. Internal validations were performed splitting the training and the test datasets in different ways. When the proteins used for training are excluded from the test dataset, the correlation coefficient between true and predicted free energies is 0.63 and most errors are false negatives. This is not a bad performance, suggesting that the model is able to generalize the training example to capture the physics of molecular recognition, but can provide too noisy predictions for applications. Testing datasets that contain a protein that is present in the training set, but very different DNA sequences, increases the correlation coefficient to 0.71. This is a typical working scenario, where one considers a specific DNA-binding protein and knows some binding sites, and want to explore the binding to different regions of the genome. Finally, the correlation coefficient of 0.93 found for a test set whose only requirement is not to contain identical pairs of proteins and DNA suggests that the model is good in predicting the effect of small mutations, either in the protein or in the DNA. As a benchmark, consider that lengthy Poisson-Boltzmann calculations, which anyway cannot be used for a high-throughput calculation of the binding profile of a protein onto a chromosome, give a correlation coefficient of 0.72[7]. Faster knoweldge-based methods used so far provide correlation coefficients lower than 0.5 and basically correlate the binding energy with the size of the DNA segment [10], something that our model does not do (Fig. S6 in the SI).

The latter case was further investigated, using the classical systems of l-repressor and Cro-repressor, for which binding experiments with radioactively-enriched DNA were done for all possible mutations[17,18]. The fact that these two proteins are dimers that binds two half-sites on DNA complicates this case. Even if this is not a purely two-state binding that also induces a bending of the double helix[28], the model can recapitulate all aspects of binding, predicting the effect of mutations.

Another validation is provided by the analysis of the distribution of energies for sites identified by ChIP-Seq experiments, which display energies that are markedly lower than those of sequences selected at random on the genome. Although ChIP-Seq is not a direct quantitative way of measuring protein binding to DNA, a qualitative correlation between the peaks in its enrichment and the binding energy is expected.

The model we developed is thus well suited for high-throughput prediction of the binding energy profile of a protein onto chromosomes. We tested the simple case of the PU.1 transcription factor, which is monomeric and binds a contiguous region of DNA. The calculation of the energy profile of murine chromosome 2 took 23 hours on a A100 gpu. The resulting energy landscape is highly roughed, supporting the kind of modelling used to explain the stability-speed paradox [27]. This roughness is not uncorrelated along the DNA. We could highlight two length scales, beyond that of variability at single base pair, one at 75 bp, which is that of genomic signals [22–25], and the other between 100 kbp and 100 Mbp, which is that of chromosomal domains [21].

An interesting feature we observed is that, although there are several low-energy sites on the chromosome, those corresponding to reported binding sites of PU.1 are in an energy basin which is wider than that of other sites with the same energy. We can speculate that this property give them a kinetic advantage, allowing the protein to reach them more easily than the others.

ASSOCIATED CONTENT

**Supporting Information**. The file supporting_information.pdf contains the extended method section and some supporting figures.

The dataset can be downloaded at https://zenodo.org/uploads/21440037

The code and the parameters can be freely downloaded at https://gitlab.com/MorfeoRenai/tf-dna-modelling

It is possible to run the code on the Colab site https://colab.research.google.com/drive/1YgB6tcpNJLFYg0UhGCyTvw_m2ShmqO7S?usp=sharing

## Supporting Information

The energy landscape of DNA-binding proteins along the genome

A. Pandolfi, R. Beccaria and G. Tiana

## Extended Methods

### ***Creation of a dataset of protein-DNA complexes and their free energies***

To facilitate automated access to NAKB, we developed *python-nakb*, a Python package that programmatically constructs and sends queries to the underlying Apache Solr search engine. To extract nucleic acid structural data, descriptors, and features from PDB structures, we developed *PDBNucleicAcids*, a Biopython-based Python package. To extract protein–DNA contacts and identify interfaces from PDB structures, we developed *biointerface*, a Python package based on *PDBNucleicAcids* itself.

Data queried from ProNAB was extensively preprocessed. Protein sequences, nucleic acid sequences, protein mutations, and nucleic acid mutations are cleaned of formatting errors and filtered when fixing is not possible; mutations are specifically filtered; all deletions were excluded from the dataset. Outliers in experimental pH and temperature were filtered out, entries with pH within 6 and 9 and temperatures under 330K were kept in the dataset. Entries with $\Delta G = 0$ were filtered out as we consider them unfit for modelling. Nucleic acid mutations, i.e. "A11T" as in "*the adenine at the 11-th position is mutated into a thymine*", are parsed and applied to the experimental sequence. For all intents and purposes, the mutated experimental sequence is the experimental sequence used for downstream modelling. If applying the mutation to the sequence fails for any reason, failing entries are filtered out.

Protein mutations are parsed and introduced in silico into the corresponding PDB biological assemblies, by replacing the side chain of target residue with the side chain of the mutated residue; we programmatically superimpose the backbone atoms of target and mutation residue, then transform the new side chain accordingly by multiplying the side chain atomic three-dimensional coordinates with the appropriate roto-translation matrix.

The codebase can be retrieved at the Gitlab repository (https://gitlab.com/MorfeoRenai/tf-dna-modelling).

### ***Machine-learning based potential***

*Model overview.* The framework designed for predicting protein-DNA binding affinity $\Delta G$ relies on a multi-modal, asymmetric dual-stream architecture optimized via end-to-end training. To separate the structural dependencies of the target protein interface from the sequential properties of the interacting DNA, the model extracts independent latent features for each entity. The physical and geometric layout of the protein binding site is processed using a continuous-filter Graph Neural Network (GNN) based on the SchNet architecture. Simultaneously, the corresponding DNA sequence is handled through a one-dimensional Convolutional Neural Network (CNN). The individual embeddings generated by these parallel streams are subsequently concatenated and passed through a Feed-Forward Neural Network (FFNN) equipped with regularized dropout layers to output a scalar estimate of $\Delta G$ .

*Protein interface graph construction.* The binding interface of the target protein is represented as a spatial geometric graph $G_p = (V_p, E_p)$. Nodes $v_i \in V_p$ correspond to individual protein atoms, while undirected edges $e_{ij} \in E_p$ capture neighborhood interactions established between atom pairs whose Euclidean distance $d_{ij}$ falls within a specified radius cutoff. During training, the binding site boundaries are defined using structural complexes from experimental datasets. Specifically, a local pocket graph is isolated by selecting all protein atoms located within a 5.5 Ang threshold of any atom in the native, co-crystallized DNA molecule.

*DNA sequence encoding.* Unlike the 3D graph representation of the protein interface, the DNA molecule is processed as a one-dimensional sequence of length $L$ . The raw nucleotide sequence is first mapped to a discrete representation using a one-hot encoding scheme, converting each base into a 4-dimensional vector representing the canonical nucleotides (A, C, G, T)

$$\mathrm{x}_{\mathrm{DNA}} \in \{0,1\}^{L\times 4}$$

To maintain consistent tensor operations across varying sequence lengths in mini-batches, shorter sequences are padded up to the maximum sequence length (max_seq_len) with zeroed vectors.

*Protein Graph Embedding through invariant SchNet GNN*

Each atom $i \in V_p$ is mapped from its atomic number z to an initial continuous hidden feature vector $\mathrm{h}_i^{(0)} \in R^d$ utilizing a learnable lookup matrix. Interatomic spatial distances $d_{ij}$ are projected into continuous edge attributes $e_{ij} \in R^K$ using a collection of $K$ Gaussian Radial Basis Functions (RBFs)

$$e_{ij} = \exp\left(-\gamma\left(d_{ij} - \mu_k\right)^2\right)$$

where $\mu_k$ samples the continuous range up to $r_{cut}$. The initial atomic embeddings are iteratively refined across $T$ sequential interaction blocks. Within each interaction block, node features undergo continuous-filter convolutions where structural features are aggregated from local spatial neighbors $\partial(i)$

$$\mathrm{x}_i' = \sum_{j\in\partial(i)\,\mathrm{x}_j} f_{\mathrm{MLP}}(\mathrm{e}_{ij}) \cdot C(d_{ij})$$

Here, $\odot$ signifies the Hadamard product, $f_{\mathrm{MLP}}$ represents a shallow feature expansion multi-layer perceptron, and $C(d_{ij})$ is a cosine cutoff function that dampens interaction strengths near the radius boundary. Following a residual update step, the final atomic embeddings are flattened through a global sum pooling operation, yielding a single, fixed-size invariant latent vector for the entire interface

$$\mathrm{o}_p = \sum_{i\in V_p}^{\mathrm{h}_i^{(T)}} \mathbb{R}^{d_{\mathrm{latent}}}$$

*DNA convolution stream.* The padded one-hot encoded matrix is transformed to match the expected shape for spatial convolutions, $\mathbb{R}^{4\times \mathrm{max_seq_len}}$, and fed into a 1D CNN stream consisting of $N_{\mathrm{layers}}$convolutional layers. The primary layer extracts base combinations using a predefined kernel size, followed by standard Rectified Linear Unit (ReLU) activations and 1D Batch

Normalization (BatchNorm1d) to stabilize optimization. Subsequent hidden convolutional layers continue to process localized contextual sequence windows

$$\mathrm{H}_{\mathrm{CNN}}^{(l)} = \mathrm{BatchNorm1d}\left(\mathrm{ReLU}\left(\mathrm{Conv1d}\left(\mathrm{H}_{\mathrm{CNN}}^{(l-1)}\right)\right)\right)$$

To collapse the spatial dimensions while remaining resilient to variable sequence lengths, an adaptive average pooling layer computes the mean along the sequence axis, generating a dense representation vector

$$\mathrm{o}_d \in \mathbb{R}^{C_{out}}$$

*Binding affinity prediction.* The structural vector of the protein interface ($\mathrm{o}_p$) and the sequence vector of the DNA ($\mathrm{o}_d$) are combined via vector concatenation to establish a unified cross-modal embedding

$$\mathrm{z}_{\mathrm{joint}}[\mathrm{o}_p| |\mathrm{o}_d]$$

This concatenated vector is passed to a non-linear Feed-Forward Neural Network (FFNN). Each intermediate layer applies a linear transformation followed by an activation function—such as the Exponential Linear Unit (ELU)—and regularized dropout to prevent over-fitting during optimization

$$\mathrm{h}_{\mathrm{FFNN}}^{(l)} = \mathrm{Dropout}\left(\mathrm{ELU}\left(\mathrm{W}^{(l)\mathrm{h}_{\mathrm{FFNN}}^{(l-1)}} + \mathrm{b}^{(l)}\right)\right)$$

The final layer projects the final dense state down to a single scalar target $\hat{y}$. To ensure the model respects the bidirectional nature of double-stranded DNA, predictions are constrained to be invariant with respect to the forward sequence and its corresponding reverse complement. For a given protein-DNA pair, the forward sequence ($\mathrm{x}_{\mathrm{DNA}}$) yields a prediction $\widehat{y_{\mathrm{fwd}}}$ . The sequence tensor is subsequently flipped along the sequence axis and inverted across the nucleotide one-hot channels to form the reverse complement ($\mathrm{x}_{\mathrm{DNA,\ rc}}$), yielding $\widehat{y_{rc}}$. The final predicted binding energy is computed as the average of both forward and reverse passes

$$\widehat{y_{final}} = \frac{1}{2}(\widehat{y_{\mathrm{fwd}}} + \widehat{y_{rc}})$$

The complete architecture is trained end-to-end. Learnable parameters are updated by minimizing the Mean Squared Error (MSE) loss between the predicted energy $\widehat{y_i}$ and the experimental value $y_i$ over a batch size$B$

$$\mathcal{L} = \frac{1}{B}\sum_{i=1}^{B}\left(\widehat{y_{i,final}} - y_i\right)^2$$

Parameter updates are governed by the Adam optimizer using adaptive moment estimation to dynamically scale gradients across both the geometric graph and the sequential convolutional streams.

*Synthetic decoy pairs.* To ensure the model learns true binding specificity rather than generic structural or sequence biases, we introduced a dynamic decoy sampling strategy during training. Training solely on native, co-crystallized complexes risks data bias, where the model pairs entities based on individual composition rather than explicit structural complementarity. To mitigate this, we augmented each true complex with $N_{\text{decoys}}$synthetic mismatched pairs by randomly assigning non-native protein interfaces to DNA sequences. Because these mismatched pairs represent non-specific binding events, their artificial affinities ($\Delta G_{\text{decoy}}$) are modeled as weak interactions dynamically sampled from a Gaussian distribution:

$$\Delta G_{\text{decoy}} \sim N(\mu = -1.0, \sigma = 0.5)$$

To prevent synthetic values from overlapping with true high-affinity binders, the sampled values are bounded using a lower threshold ($\tau = -2.5$ ):

$$\Delta G_{decoy,final} = \max\left(\Delta G_{\text{decoy}}, \tau\right)$$

To maintain a balanced optimization signal, sample loss contributions are regularized via scaling weights ($w$ ). True complexes receive a full weight ($w_{\text{true}} = 1$), while individual decoys are downscaled proportionally ($w_{\text{decoy}} = \frac{1}{N_{\text{decoy}}}$). This penalization framework balances the aggregate gradient impact of synthetic pairs against native data, preventing model collapse and forcing the GNN-CNN architecture to capture true, sequence-dependent interface specificity.

## Supporting Figures

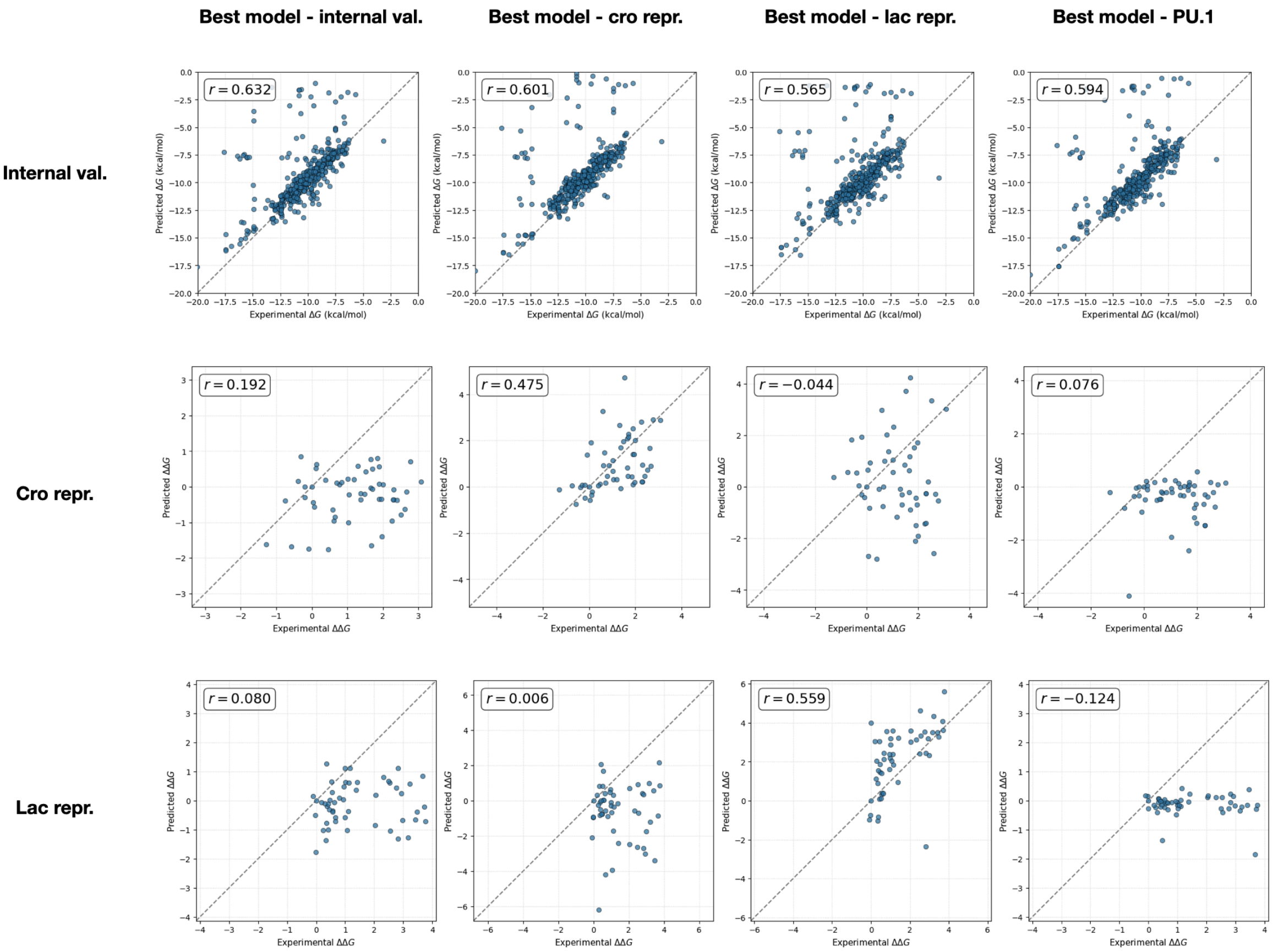

Figure S1: The Pearson correlation between experimental and predicted data for the internal validation and for the calculation of $\Delta\Delta G$ with the models optimized for each of the two tasks.

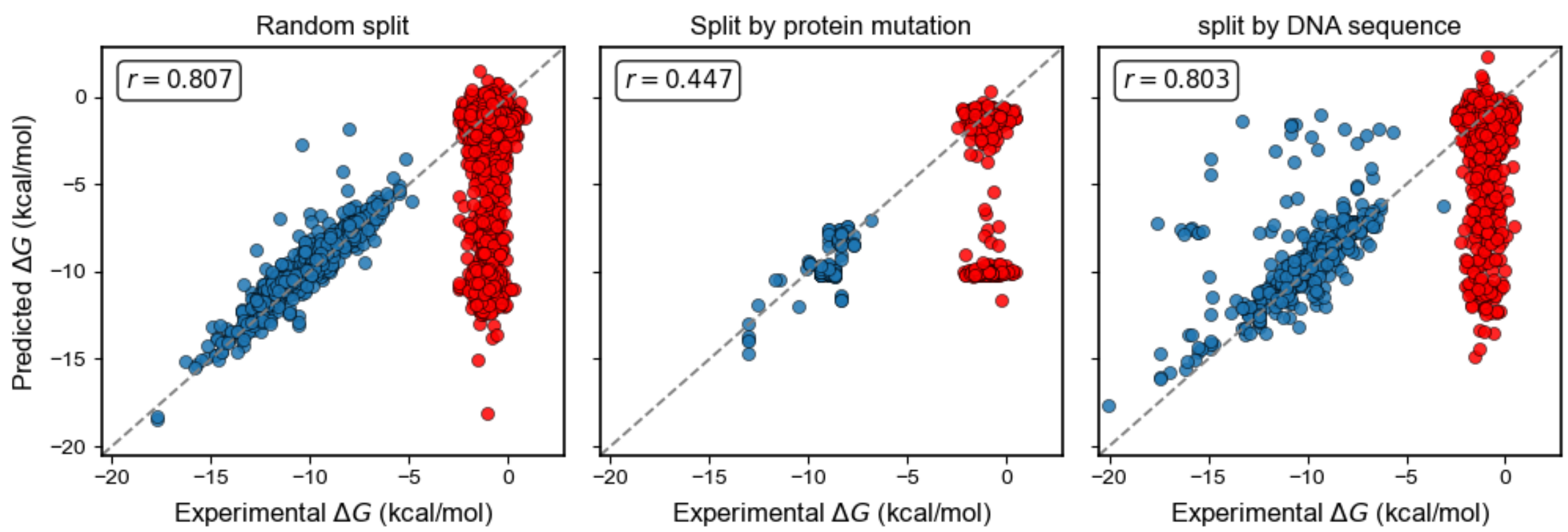

Figure S2: same plot as Fig 3 in main manuscript but including decoys in scatterplot (red) and calculations of Pearson coefficients.

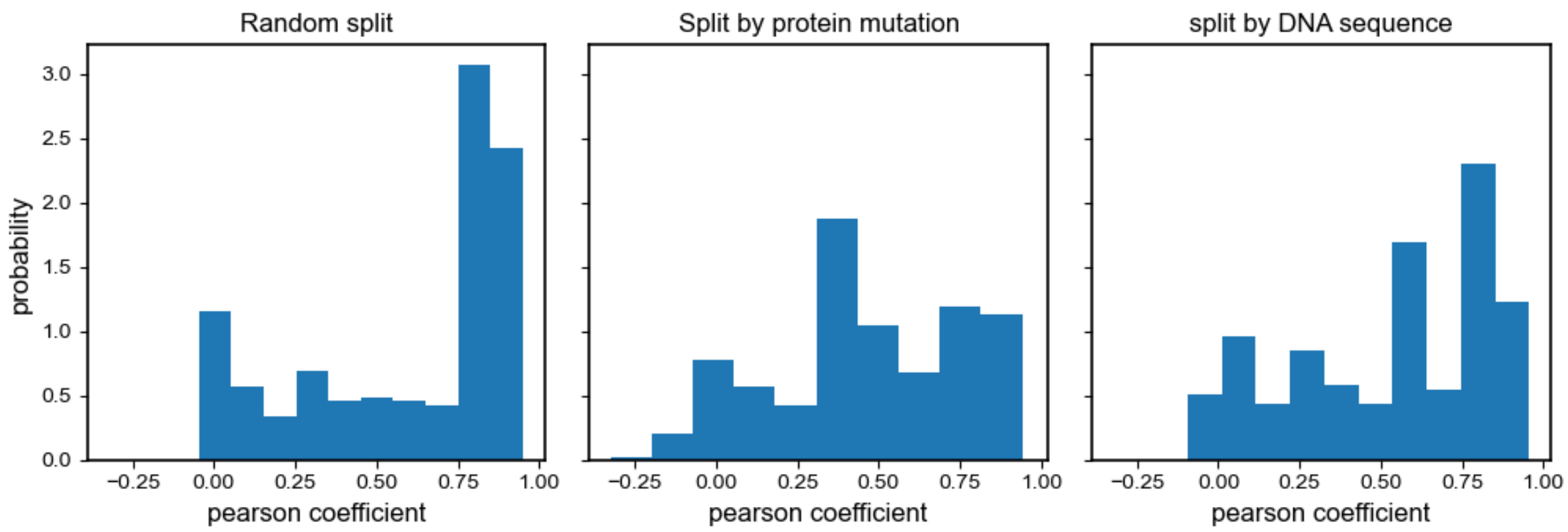

Figure S3: Distribution of Pearson coefficient values on validation set for all trainings (i.e. different hyperparameter values) across the three different splitting strategies.

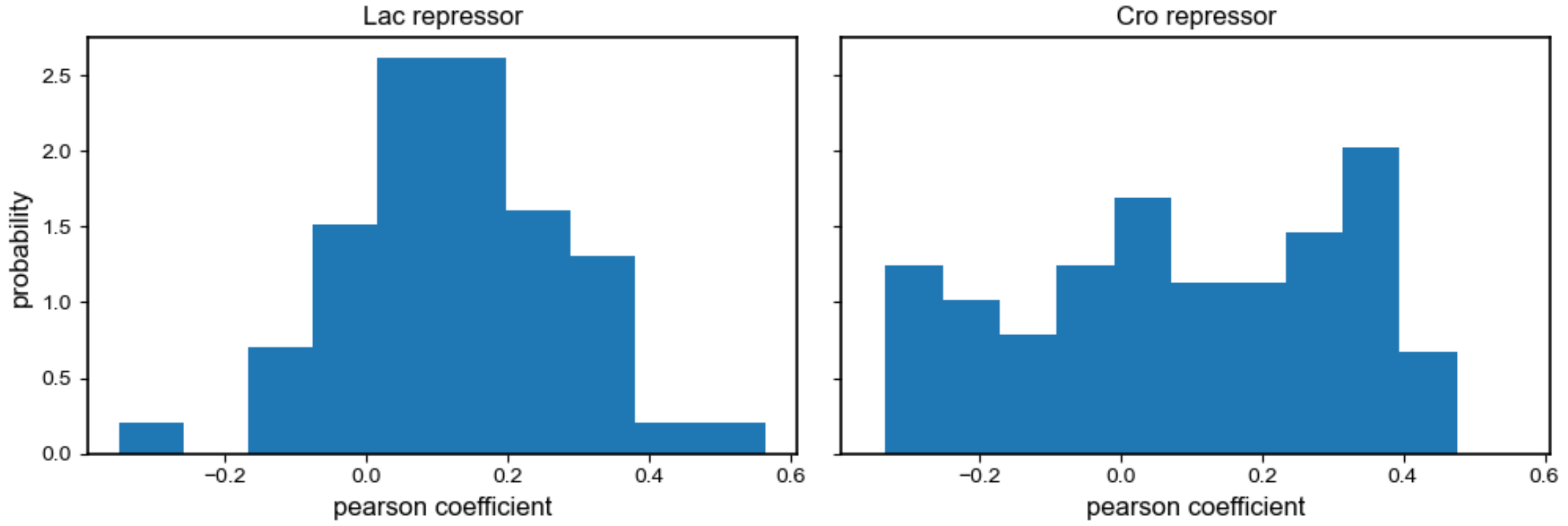

Figure S4: Distribution of Pearson coefficient values on Lac and Cro validation for all models trained using DNA similarity splitting strategy.

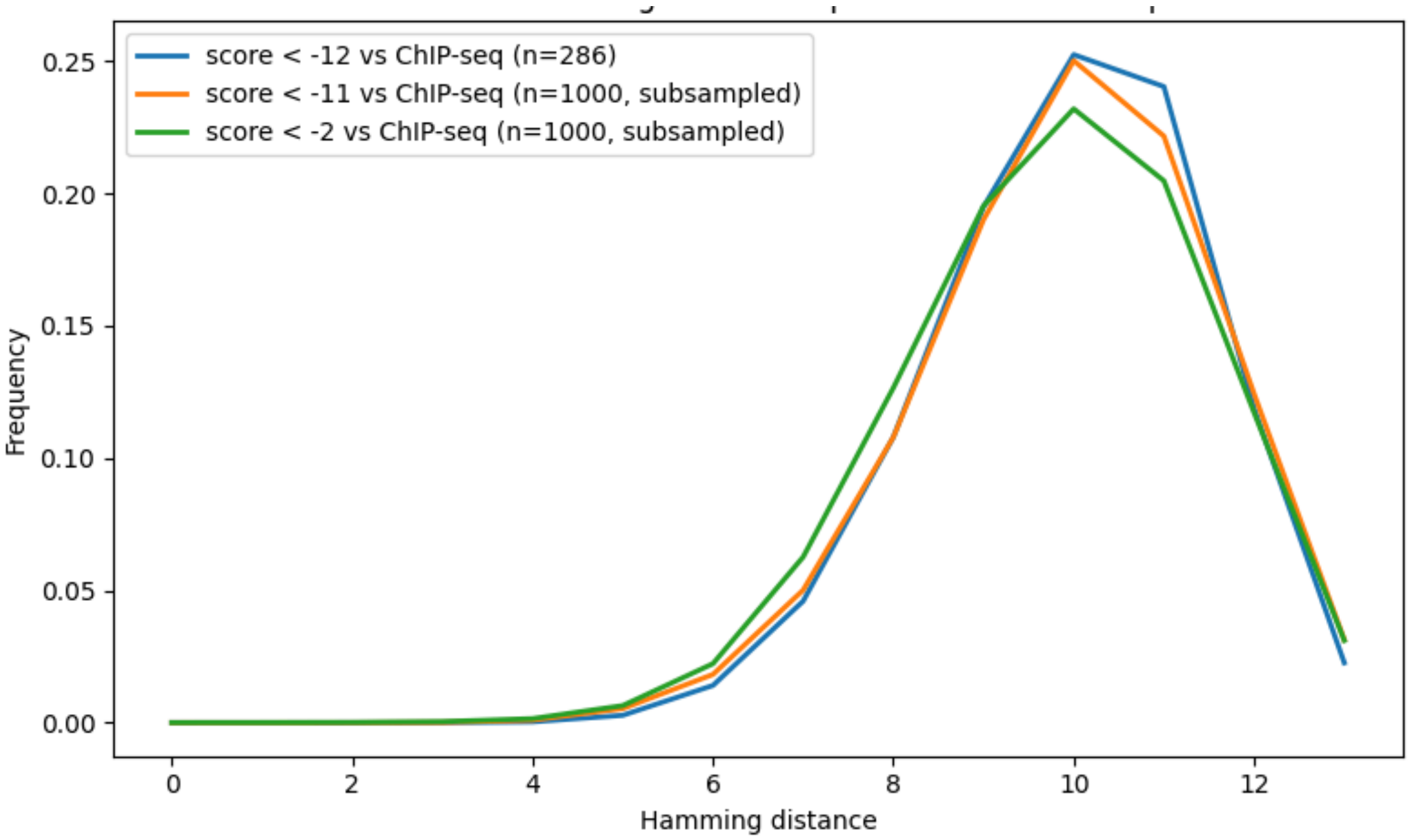

Figure S5: The distribution of Hamming distances between pairs of binding sequences of chromosome 2 with energy below -12 and those obtained in chip-seq experiments (blue curve). The same, using chromosomal sequences with energy lower than -11 (orange curve) and -2 (green curve).

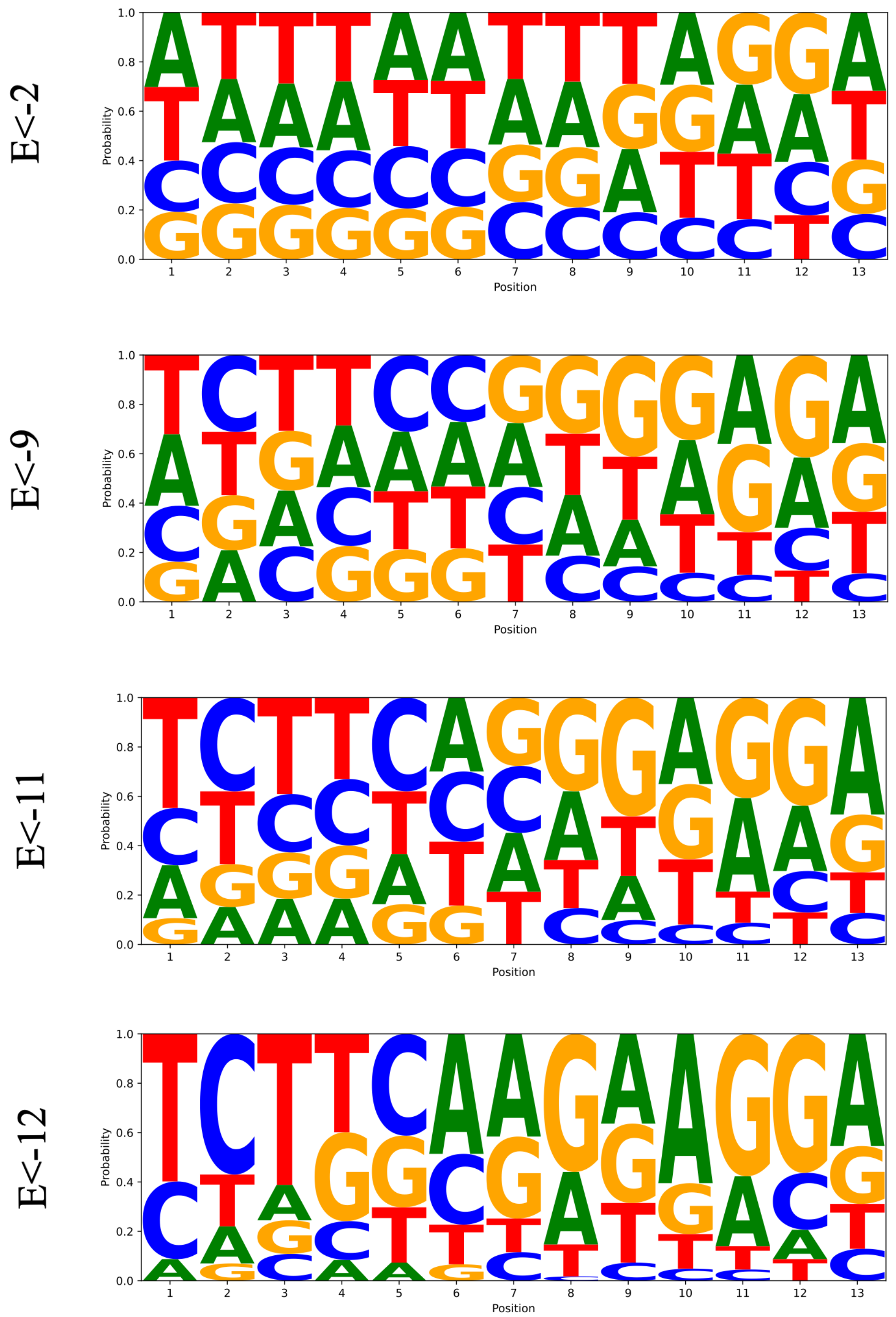

Figure S6: The probability of bases in sites whose binding energy is predicted to be lower than -2, -9, -11 and -12 kcal/mol, respectively.

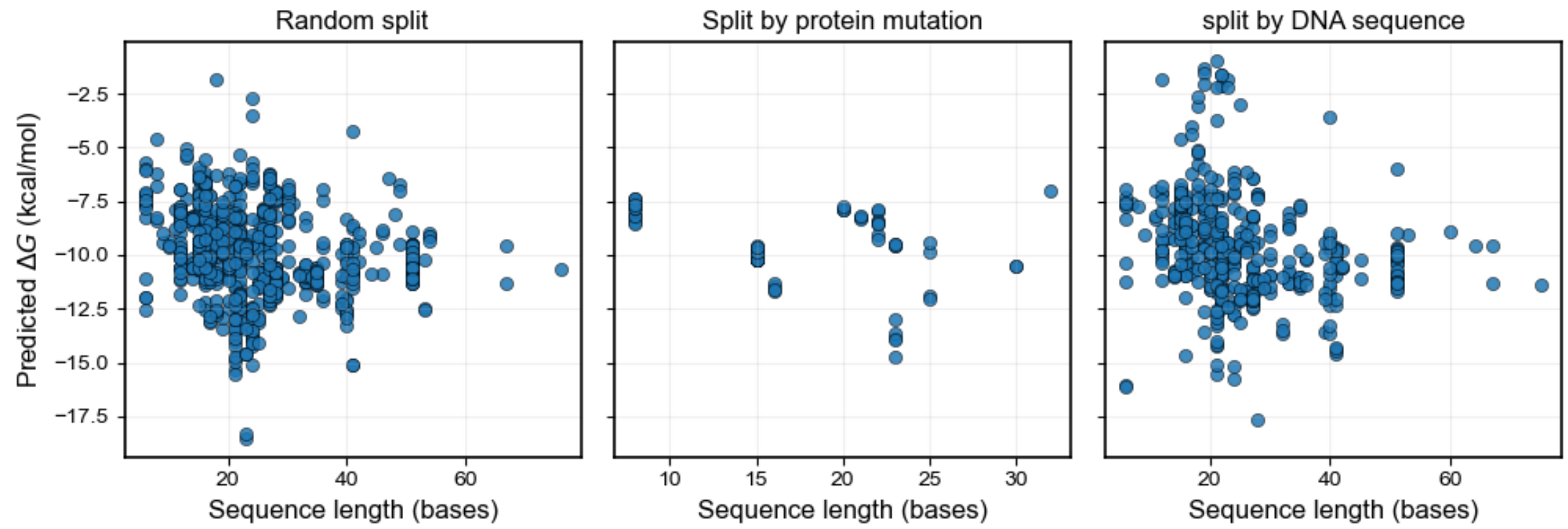

Figure S7: No trivial correlation with number of contacts

| name | Internal val | Lac rep. | Cro rep. | PU.1 | description |
|---|---|---|---|---|---|
| num_epochs | 50 | 50 | 50 | 50 | Maximum training epochs |
| batch_size | 128 | 128 | 128 | 128 | Samples per batch |
| seed | 2026 | 2026 | 2026 | 2026 | Random seed |
| num_decoys | 3 | 3 | 3 | 3 | Decoys per positive sample |
| decoy_mu | −1.0 | −1.0 | −1.0 | −1.0 | Decoy $\Delta G$ mean |
| decoy_sigma | 0.5 | 0.5 | 0.5 | 0.5 | Decoy $\Delta G$ standard deviation |
| decoy_threshold | −2.5 | −2.5 | −2.5 | −2.5 | Minimum decoy $\Delta G$ |
| schnet_hidden_channels | 128 | 128 | 128 | 128 | SchNet hidden dimension |
| schnet_latent_channels | 128 | 128 | 256 | 128 | SchNet output dimension |
| schnet_num_filters | 128 | 128 | 128 | 128 | SchNet filter dimension |
| schnet_num_interactions | 6 | 6 | 6 | 6 | SchNet interaction blocks |
| schnet_num_gaussians | 50 | 50 | 50 | 50 | Gaussian basis functions |
| schnet_cutoff | 3.5 | 3.5 | 3.5 | 3.5 | Neighbor cutoff distance (Ang) |
| cnn_num_layers | 2 | 3 | 3 | 3 | CNN layers |
| cnn_channels | 128 | 64 | 64 | 128 | CNN channels |
| cnn_kernel_size | 5 | 5 | 5 | 5 | CNN kernel size |
| ffnn_num_layers | 2 | 3 | 2 | 2 | FFNN hidden layers |
| ffnn_num_nodes | 128 | 256 | 256 | 128 | FFNN hidden nodes |
| ffnn_activation_func | ELU | ELU | ELU | ELU | FFNN activation |
| ffnn_dropout_p | 0.2 | 0.2 | 0.2 | 0.2 | Dropout probability |
| weight_decay | 0.0 | 0.0 | 0.0 | 0.0 | L2 regularization for optimizer |
| learning_rate | 0.001 | 0.001 | 0.001 | 0.001 | Optimizer learning rate |
| early_stopping_patience | 5.0 | 5.0 | 5.0 | 5.0 | Early stopping patience |

Table S1: The optimal set of hyperparameters for the internal validation and for the calculation of $\Delta\Delta G$.